\documentclass[aps,preprint,showpacs,preprintnumbers,amsmath,amssymb]{revtex4}
\usepackage{amsmath,mathrsfs,amsbsy,color,graphicx,bm,amsthm,amsfonts}
\usepackage{units}
\usepackage{bbm}
\usepackage{times}
\usepackage{dcolumn}
\usepackage{mathrsfs}% Align table columns on decimal point
\usepackage{amsmath,amssymb,epsfig}
\usepackage{amsmath}
\newcommand{\udots}{\mathinner{\mskip1mu\raise1pt\vbox{\kern7pt\hbox{.}}
\mskip2mu\raise4pt\hbox{.}\mskip2mu\raise7pt\hbox{.}\mskip1mu}}
\begin{document}
\title{Freezing of maximal negativity for a four-qubit cluster state in Schwarzschild spacetime}
\author{Si-Han Li, Hui-Chen Yang, Rui-Yang Xu, Shu-Min Wu\footnote{Email: smwu@lnnu.edu.cn}}
\affiliation{ Department of Physics, Liaoning Normal University, Dalian 116029, China}

% \baselineskip=0.65 cm

%\vspace*{0.2cm}
\begin{abstract}
We demonstrate exact freezing of maximal bipartite negativity for a four-qubit cluster state of fermionic modes in Schwarzschild spacetime. Within a single-mode description, the Hawking transformation is applied to one mode, $D$, and its inaccessible interior partner is traced out. The negativities across $A|BCD$ and $B|ACD$ remain exactly equal to one at all Hawking temperatures, while those across the other two single-qubit-versus-rest bipartitions decrease. An analytical factorization exposes an unaffected Bell-state factor and explains this invariance. Applying the transformation to different modes identifies the corresponding invariant bipartitions, while comparisons with four-qubit GHZ and W states reveal distinct responses under the same map. The Bell-factor argument further establishes invariance under any completely positive trace-preserving map confined to $D$, showing that the preservation mechanism is independent of the particular thermal coefficients.
\end{abstract}

\vspace*{0.5cm}
\maketitle
\section{Introduction}
Quantum information in curved spacetime lies at the intersection of quantum information theory, quantum field theory, and general relativity, aiming to reveal how gravity influences fundamental quantum resources such as entanglement and coherence, to explore methods for preserving these resources in relativistic settings, and to utilize quantum technologies for probing spacetime structure \cite{SDF1,SDF2,SDF3,SDF4,SDF5,SDF6,SDF7,SDF8,SDF9,SDF10,SDF11,SDF12,SDF13,SDF14,SDF15,SDF16,SDF17,SDF18,SDF19,SDF20,SDF21,SDF22,SDF23,SDF24,SDF25,SDF26,SDF27,SDF28,SDF29,SDF30,SDF31,SDF32,SDF33,SDF34,SDF35,SDF36,SDF37,SDF38,SDF39,SDF40,SDF41,SDF42,SDF43,SDF44,SDF45,SDF46,SDF47,SDF48,SDF49,SDF50,SDF51,
SDF52,SDF53,SDF54,SDF55,QBM1,QBM2,QBM4,QBM5,QBM6,QBM7,QBM8}. A central research direction in this field involves studying quantum systems within the single-mode approximation of free field models.
Extensive studies have demonstrated entanglement degradation in a variety of settings involving the Unruh and Hawking effects. Representative examples include Bell states~\cite{SDF3,SDF4,SDF8} as well as multipartite GHZ and W states~\cite{AGH1,AGH2,AGH3,AGH4,AGH5,AGH6,AGH7}. Relativistic effects can also redistribute, generate, or enhance correlations, depending on the state, the selected modes, the accessible degrees of freedom, and the bipartition~\cite{SDF39,SDF49}. This diversity motivates the identification of states and mode configurations that preserve entanglement under a given relativistic transformation.

As relativistic quantum information tasks become more complex, multipartite entanglement plays a crucial role in supporting advanced quantum protocols. Identifying quantum states that maintain their quantum features in extreme gravitational conditions has thus become a pressing research goal. Cluster states, particularly the four-qubit $CL_4$ state, represent a unique and valuable class of quantum resources \cite{LSH1,LSH2,LSH3,LSH4,LSH5,LSH6,LSH7,LSH8,LSH9,LSH10,LSH11,LSH12}.
Their graph-theoretic structure underlies measurement-based quantum computation and has motivated extensive studies of entanglement robustness under local noise~\cite{LSH8,LSH9,LSH10,LSH11,LSH12,LSH13,LSH14,LSH15,LSH16}. A natural question is whether particular bipartitions of the four-qubit cluster state can retain maximal negativity under a fermionic Hawking transformation. Establishing the mechanism of such freezing would clarify how the entanglement structure of the input state and the location of the affected mode govern the persistence of bipartite correlations.

General relativity predicts the potential existence of black holes, a prediction that has since been extensively developed in the context of black hole physics and corroborated by both indirect and direct astronomical observations. Motivated by these advances, we systematically investigate the properties of quantum entanglement of the $CL_4$ state in the curved spacetime of a Schwarzschild black hole.
We consider four fermionic mode subsystems and apply the Hawking transformation to mode $D$. The cluster state exhibits exact freezing of maximal negativity across $A|BCD$ and $B|ACD$, while its other two single-qubit-versus-rest negativities decrease with the Hawking temperature. We derive an analytical explanation by exposing a Bell-state factor unaffected by the transformation. Applying the same map to the other modes determines which bipartitions remain invariant, and comparisons with GHZ and W states illustrate the dependence on the input entanglement structure. The resulting factorization also establishes invariance under arbitrary completely positive trace-preserving maps confined to $D$, within the same mode description.

The paper is organized as follows. In Sec.~II, we introduce the quantization of the Dirac field in the  Schwarzschild black hole background. In Sec.~III, we analyze the entanglement properties of the $CL_4$ state in this background. Finally, Sec.~IV concludes with a summary of our findings.

\section{Quantization of Dirac field in Schwarzschild spacetime }

The metric of the Schwarzschild black hole is given by
\begin{eqnarray}\label{w11}
ds^2&=&-(1-\frac{2M}{r}) dt^2+(1-\frac{2M}{r})^{-1} dr^2\nonumber\\&&+r^2(d\theta^2
+\sin^2\theta d\varphi^2).
\end{eqnarray}
Here, $r$ is the Schwarzschild radial coordinate, $M$ is the black-hole mass, and the event horizon is located at $r=2M$ \cite{SDF4}.
For simplicity, we adopt natural units where $\hbar = G = c = k = 1$. In this spacetime, the massless Dirac equation $[\gamma^a e_a{}^\mu(\partial_\mu+\Gamma_\mu)]\Phi=0$ takes the explicit form
\begin{eqnarray}\label{w12}
&&-\frac{\gamma_0}{\sqrt{1-\frac{2M}{r}}}\frac{\partial \Phi}{\partial t}+\gamma_1\sqrt{1-\frac{2M}{r}}\bigg[\frac{\partial}{\partial r}+\frac{1}{r}+\frac{M}{2r(r-2M)} \bigg]\Phi \nonumber\\
&&+\frac{\gamma_2}{r}(\frac{\partial}{\partial \theta}+\frac{\cot \theta}{2})\Phi+\frac{\gamma_3}{r\sin\theta}\frac{\partial\Phi}{\partial\varphi}=0,
\end{eqnarray}
where $\gamma_i$ ($i=0,1,2,3$) represent the Dirac matrices  \cite{SDF6,SDF12}.

We consider a fixed, eternal Schwarzschild black hole and choose the Hartle--Hawking state as the reference vacuum. This state is regular across the past and future horizons and describes thermal equilibrium between incoming and outgoing radiation. The exterior number states are defined in the Boulware basis by positive frequency with respect to Schwarzschild Killing time. The calculation retains one outgoing fermionic mode sector and its interior antiparticle partner, with the remaining independent sectors and spin labels suppressed. Both incoming and outgoing sectors obey the Hartle--Hawking boundary conditions; the expansions below display the retained outgoing sector. This equilibrium setting is distinct from the Unruh state appropriate to the late-time radiation of a collapsing black hole~\cite{Boulware1975,HartleHawking1976,Unruh1976}.

Solving the near-horizon Dirac equation gives the selected outgoing solutions inside and outside the horizon, respectively:
\begin{eqnarray}\label{w13}
\Phi^+_{{\mathbf{k}},{\rm in}}\sim \phi(r) e^{i\omega u},
\end{eqnarray}
\begin{eqnarray}\label{w14}
\Phi^+_{{\mathbf{k}},{\rm out}}\sim \phi(r) e^{-i\omega u},
\end{eqnarray}
where $\phi(r)$ is the four-component Dirac spinor and $u=t-r_{*}$ is the retarded time, with the tortoise coordinate defined as $r_{*}=r+2M\ln\frac{r-2M}{2M}$ \cite{SDF22}.
Here $\omega>0$ is the exterior mode frequency with respect to Schwarzschild Killing time $t$, and $\mathbf{k}$ labels the selected mode. The retained outgoing sector admits the Schwarzschild-mode expansion
\begin{eqnarray}\label{w15}
\Phi&=&\int
d\mathbf{k}[\hat{a}^{\rm in}_{\mathbf{k}}\Phi^{+}_{{\mathbf{k}},\text{in}}
+\hat{b}^{\rm in \dagger}_{\mathbf{k}}
\Phi^{-}_{{\mathbf{k}},\text{in}}\nonumber\\ &+&\hat{a}^{\rm out}_{\mathbf{k}}\Phi^{+}_{{\mathbf{k}},\text{out}}
+\hat{b}^{\rm out \dagger}_{\mathbf{k}}\Phi^{-}_{{\mathbf{k}},\text{out}}],
\end{eqnarray}
where $\hat{a}^{\rm in}_{\mathbf{k}}$ and $\hat{b}^{\rm in\dagger}_{\mathbf{k}}$ are the fermion annihilation and antifermion creation operators inside the event horizon, while $\hat{a}^{\rm out}_{\mathbf{k}}$ and $\hat{b}^{\rm out\dagger}_{\mathbf{k}}$ are the corresponding operators for the exterior region. These operators satisfy the canonical anticommutation relations $\{\hat{a}^{\rm out}_{\mathbf{k}},\hat{a}^{\rm out\dagger}_{\mathbf{k'}}\}=
\{\hat{b}^{\rm in}_{\mathbf{k}},\hat{b}^{\rm in\dagger}_{\mathbf{k'}}\}
=\delta_{\mathbf{k}\mathbf{k'}}.$The exterior Schwarzschild number vacuum is the Boulware vacuum, satisfying $\hat a_{\mathbf{k}}^{\mathrm{out}}|0\rangle_B=\hat b_{\mathbf{k}}^{\mathrm{out}}|0\rangle_B=0$; the interior kets below label partner occupation numbers in the near-horizon decomposition.

Following Damour and Ruffini~\cite{AAAA3}, analytic continuation of Eqs.~(\ref{w13}) and~(\ref{w14}) across the horizon yields the horizon-analytic modes for the retained outgoing sector:
\begin{eqnarray}\label{w16}
\Psi^+_{{\mathbf{k}},{\rm out}}=e^{-2\pi M\omega} \Phi^-_{{-\mathbf{k}},{\rm in}}+e^{2\pi M\omega}\Phi^+_{{\mathbf{k}},{\rm out}},
\end{eqnarray}
\begin{eqnarray}\label{w17}
\Psi^+_{{\mathbf{k}},{\rm in}}=e^{-2\pi M\omega} \Phi^-_{{-\mathbf{k}},{\rm out}}+e^{2\pi M\omega}\Phi^+_{{\mathbf{k}},{\rm in}}.
\end{eqnarray}
The same outgoing sector can be expanded in these horizon-analytic modes as
\begin{eqnarray}\label{w18}
\Phi&=&\int
d\mathbf{k} [2\cosh(4\pi M\omega)]^{-\frac{1}{2}}
[\hat{c}^{\rm in}_{\mathbf{k}}\Psi^{+}_{{\mathbf{k}},\text{in}}
+\hat{d}^{\rm in\dagger}_{\mathbf{k}}
\Psi^{-}_{{\mathbf{k}},\text{in}}\nonumber\\ &+&\hat{c}^{\rm out}_{\mathbf{k}}\Psi^{+}_{{\mathbf{k}},\text{out}}
+\hat{d}^{\rm out\dagger}_{\mathbf{k}}\Psi^{-}_{{\mathbf{k}},\text{out}}].
\end{eqnarray}
Here, $\hat c_{\mathbf{k}}^\sigma$ and $\hat d_{\mathbf{k}}^\sigma$, with $\sigma=(\mathrm{in},\mathrm{out})$, are the particle and antiparticle annihilation operators in the horizon-analytic basis. The notation $|0\rangle_K$ denotes the selected two-mode factor of the Hartle--Hawking reference vacuum.

Comparing the expansions in Eqs.(\ref{w15}) and (\ref{w18}) yields the Bogoliubov transformation relating the Schwarzschild and Kruskal operators  \cite{SDF14}
\begin{eqnarray}\label{w19}
\hat{c}^{\rm out}_{\mathbf{k}}=\frac{1}{\sqrt{e^{-8\pi M\omega}+1}}\hat{a}^{\rm out}_{\mathbf{k}}-\frac{1}{\sqrt{e^{8\pi M\omega}+1}}\hat{b}^{\rm in\dagger}_{\mathbf{k}},
\end{eqnarray}
\begin{eqnarray}\label{ww19}
\hat{c}^{\rm out\dagger}_{\mathbf{k}}=\frac{1}{\sqrt{e^{-8\pi M\omega}+1}}\hat{a}^{\rm out\dagger}_{\mathbf{k}}-\frac{1}{\sqrt{e^{8\pi M\omega}+1}}\hat{b}^{\rm in}_{\mathbf{k}}.
\end{eqnarray}
For the selected sector, the reference vacuum and one-particle excitation are related to the exterior and interior number states by
\begin{eqnarray}\label{w20}
\nonumber |0\rangle_K&=&\frac{1}{\sqrt{e^{-\frac{\omega}{T}}+1}}|0\rangle_{\rm out} |0\rangle_{\rm in}+\frac{1}{\sqrt{e^{\frac{\omega}{T}}+1}}|1\rangle_{\rm out} |1\rangle_{\rm in},\\
|1\rangle_K&=&|1\rangle_{\rm out} |0\rangle_{\rm in}.
\end{eqnarray}
Here, $T\equiv T_H=(8\pi M)^{-1}$ is the Hawking temperature, and $\omega$ is the Schwarzschild Killing frequency. The states $|n\rangle_{\mathrm{out}}$ and $|n\rangle_{\mathrm{in}}$ denote exterior fermion and interior antifermion number states, respectively; ``out'' and ``in'' refer to location relative to the horizon~\cite{SDF1,AGH4}. The mean exterior occupation number is
\begin{eqnarray}\label{w21}
N_F=\sideset{_K}{}{\mathop{\langle}}0|\hat{a}^{\rm out\dagger}_{\mathbf{k}}\hat{a}^{\rm out}_{\mathbf{k}}|0\rangle_K=\frac{1}{e^{\frac{\omega}{T}}+1}.
\end{eqnarray}
Eq.(\ref{w21}) gives a Fermi--Dirac occupation number at temperature $T_H$. For a static observer at $r>2M$, proper time and the locally measured frequency and equilibrium temperature satisfy \cite{SDF7}
\[
d\tau=\sqrt{f(r)}\,dt,\qquad
\omega_{\mathrm{loc}}=\frac{\omega}{\sqrt{f(r)}},\qquad
T_{\mathrm{loc}}=\frac{T_H}{\sqrt{f(r)}},\qquad
f(r)=1-\frac{2M}{r}.
\]
Consequently, $\omega_{\mathrm{loc}}/T_{\mathrm{loc}}=\omega/T_H$ for the same Killing-frequency mode. Figs.\ref{Fig1} and \ref{Fig2} scan $T_H$ at fixed $\omega$, comparing Schwarzschild backgrounds of different masses. A static detector with a fixed local gap at different radii instead samples different Killing frequencies.

\section{Quantum entanglement for the $CL_{4}$ state in Schwarzschild spacetime}

\begin{figure}
\centering
\includegraphics[height=2.3in,width=3.5in]{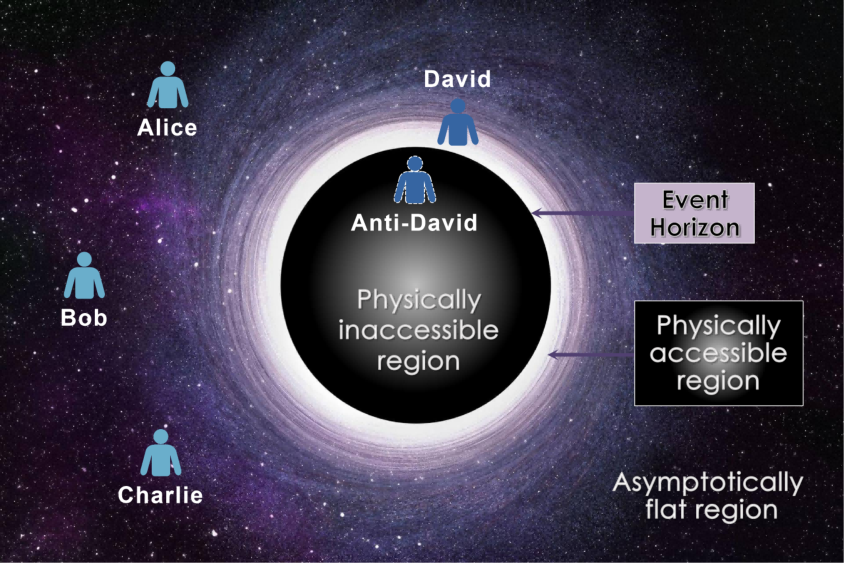}
\caption{Schematic illustration of mode accessibility. Alice, Bob, and Charlie label the reference modes $A,B,C$; David labels the retained exterior mode $D$, and $\bar D$ denotes its inaccessible interior partner. The horizon separates the retained exterior mode from the interior partner that is traced out.}
\label{Fig0}
\end{figure}

We study the four-qubit cluster state encoded in four fermionic occupation-number mode subsystems:
\begin{eqnarray}\label{w23}
|\Psi_{ABCD}^{CL_{4}}\rangle=\frac{1}{2}(|0_A0_B0_C0_D\rangle+|0_A0_B1_C1_D\rangle+|1_A1_B0_C0_D\rangle-|1_A1_B1_C1_D\rangle).
\end{eqnarray}
The subscripts $A,B,C,D$ label the mode subsystems associated with Alice, Bob, Charlie, and David, respectively. Modes $A,B,C$ serve as reference subsystems with unchanged mode descriptions. The input mode $D$ is expressed in the exterior/interior basis of Eq.~(\ref{w20}), after which $D$ denotes the retained exterior mode and $\bar D$ its inaccessible interior partner. Tracing over $\bar D$ gives the accessible four-mode state. Fig.\ref{Fig0} schematically illustrates this division of mode access.

Applying Eq.~(\ref{w20}) to mode $D$, we express the cluster state in terms of the exterior mode and its interior partner $\bar D$:
\begin{eqnarray}\label{w25}
|\Psi^{CL_{4}}_{ABCD{\bar D}}\rangle&=&
\frac{1}{2 \sqrt{e^{-\frac{\omega}{T}}+1}} \big(|0\rangle_{A}|0\rangle_{B}|0\rangle_{C}|0\rangle_{D}|0\rangle_{\bar D}+|1\rangle_{A}|1\rangle_{B}|0\rangle_{C}|0\rangle_{D}|0\rangle_{\bar D} \big)\nonumber\\&&
+\frac{1}{2 \sqrt{e^{\frac{\omega}{T}}+1}} \big(|0\rangle_{A}|0\rangle_{B}|0\rangle_{C}|1\rangle_{D}|1\rangle_{\bar D}+|1\rangle_{A}|1\rangle_{B}|0\rangle_{C}|1\rangle_{D}|1\rangle_{\bar D} \big)\nonumber\\&&
+\frac{1}{2} \big(|0\rangle_{A}|0\rangle_{B}|1\rangle_{C}|1\rangle_{D}|0\rangle_{\bar D}-|1\rangle_{A}|1\rangle_{B}|1\rangle_{C}|1\rangle_{D}|0\rangle_{\bar D} \big).
\end{eqnarray}
Here, $\bar D$ denotes the inaccessible interior partner associated with mode $D$. Tracing over this partner yields the accessible four-mode density matrix:
\begin{eqnarray}\label{w27}
 \rho^{CL_{4}}_{ABCD}=\frac{1}{4}
 \left(\!\!\begin{array}{cccccc}
\frac{1}{e^{-\frac{\omega}{T}}+1} & 0 & \frac{1}{\sqrt{e^{-\frac{\omega}{T}}+1}} & \frac{1}{e^{-\frac{\omega}{T}}+1} & 0 & -\frac{1}{\sqrt{e^{-\frac{\omega}{T}}+1}} \\
0 & \frac{1}{e^{\frac{\omega}{T}}+1} & 0 & 0 & \frac{1}{e^{\frac{\omega}{T}}+1} & 0 \\
\frac{1}{\sqrt{e^{-\frac{\omega}{T}}+1}} & 0 & 1 & \frac{1}{\sqrt{e^{-\frac{\omega}{T}}+1}} & 0 & -1 \\
\frac{1}{e^{-\frac{\omega}{T}}+1} & 0 & \frac{1}{\sqrt{e^{-\frac{\omega}{T}}+1}} & \frac{1}{e^{-\frac{\omega}{T}}+1} & 0 & -\frac{1}{\sqrt{e^{-\frac{\omega}{T}}+1}} \\
0 & \frac{1}{e^{\frac{\omega}{T}}+1} & 0 & 0 & \frac{1}{e^{\frac{\omega}{T}}+1} & 0 \\
-\frac{1}{\sqrt{e^{-\frac{\omega}{T}}+1}} & 0 & -1 & -\frac{1}{\sqrt{e^{-\frac{\omega}{T}}+1}} & 0 & 1
\end{array}\!\!\right),
\end{eqnarray}
where the matrix is expressed in the ordered basis $|0000\rangle$, $|0001\rangle$, $|0011\rangle$, $|1100\rangle$, $|1101\rangle$ and $|1111\rangle$.

To quantify the quantum entanglement between mode subsystems in this curved spacetime, we employ the negativity, a widely used entanglement measure applicable to both pure and mixed states.
It is particularly effective for detecting entanglement: a bipartitioned density matrix $\rho$ is entangled whenever its partial transpose exhibits at least one negative eigenvalue.
Specifically, for a four-partite (tetrapartite) system, the negativity associated with a given bipartition is defined as
\begin{eqnarray}\label{TH1}
N_{A(BCD)}=\|\rho^{{T_{A}}}_{ABCD}\|-1.
\end{eqnarray}
Here, $N_{A(BCD)}$ quantifies bipartite entanglement across $A|BCD$, $T_A$ denotes partial transposition with respect to $A$, and $\|\rho^{{T_{A}}}_{ABCD}\|$ denotes the trace norm of the partially transposed density matrix~\cite{NE1}.
With the normalization in Eq.~(\ref{TH1}), the upper bound for each single-qubit-versus-rest negativity is one. We refer to saturation of this bound as maximal negativity.
Alternatively, using the trace-norm definition $\|O\|=\operatorname{Tr}\sqrt{O^{\dagger}O}$ for a Hermitian operator $O$, the negativity can be rewritten in a computationally more convenient form
\begin{eqnarray}\label{TH2}
\|M\|-1=2\sum^{N}_{i=1}|\lambda^{(-)}_{M}|^{i},
\end{eqnarray}
where $|\lambda^{(-)}_{M}|^{i}$ denotes the absolute values of the negative eigenvalues of the partially transposed matrix $M$. This expression provides a direct and numerically efficient way to evaluate the entanglement for a given bipartition.

We now evaluate the single-qubit-versus-rest negativities of $\rho^{CL_4}_{ABCD}$. Taking the partial transpose with respect to $A$, we obtain
\begin{eqnarray}\label{TH3}
\scalebox{0.71}{$
\rho^{CL_{4}({T_{A}})}_{ABCD}=
\frac{1}{4}
 \left(\!\!\begin{array}{cccccccccccc}
\frac{1}{e^{-\frac{\omega}{T}}+1} & 0 & \frac{1}{\sqrt{e^{-\frac{\omega}{T}}+1}} & 0 & 0 & 0 & 0 & 0 & 0 & 0 & 0 & 0 \\
0 & \frac{1}{e^{\frac{\omega}{T}}+1} & 0 & 0 & 0 & 0 & 0 & 0 & 0 & 0 & 0 & 0 \\
\frac{1}{\sqrt{e^{-\frac{\omega}{T}}+1}} & 0 & 1 & 0 & 0 & 0 & 0 & 0 & 0 & 0 & 0 & 0 \\
0 & 0 & 0 & 0 & 0 & 0 & \frac{1}{e^{-\frac{\omega}{T}}+1} & 0 & \frac{1}{\sqrt{e^{-\frac{\omega}{T}}+1}} & 0 & 0 & 0 \\
0 & 0 & 0 & 0 & 0 & 0 & 0 & \frac{1}{e^{\frac{\omega}{T}}+1} & 0 & 0 & 0 & 0 \\
0 & 0 & 0 & 0 & 0 & 0 & -\frac{1}{\sqrt{e^{-\frac{\omega}{T}}+1}} & 0 & -1 & 0 & 0 & 0 \\
0 & 0 & 0 & \frac{1}{e^{-\frac{\omega}{T}}+1} & 0 & -\frac{1}{\sqrt{e^{-\frac{\omega}{T}}+1}} & 0 & 0 & 0 & 0 & 0 & 0 \\
0 & 0 & 0 & 0 & \frac{1}{e^{\frac{\omega}{T}}+1} & 0 & 0 & 0 & 0 & 0 & 0 & 0 \\
0 & 0 & 0 & \frac{1}{\sqrt{e^{-\frac{\omega}{T}}+1}} & 0 & -1 & 0 & 0 & 0 & 0 & 0 & 0 \\
0 & 0 & 0 & 0 & 0 & 0 & 0 & 0 & 0 & \frac{1}{e^{-\frac{\omega}{T}}+1} & 0 & -\frac{1}{\sqrt{e^{-\frac{\omega}{T}}+1}} \\
0 & 0 & 0 & 0 & 0 & 0 & 0 & 0 & 0 & 0 & \frac{1}{e^{\frac{\omega}{T}}+1} & 0 \\
0 & 0 & 0 & 0 & 0 & 0 & 0 & 0 & 0 & -\frac{1}{\sqrt{e^{-\frac{\omega}{T}}+1}} & 0 & 1 \\
\end{array}\!\!\right)$},
\end{eqnarray}
where the matrix is expressed in the ordered basis $|0000\rangle$, $|0001\rangle$, $|0011\rangle$, $|0100\rangle$, $|0101\rangle$, $|0111\rangle$, $|1000\rangle$, $|1001\rangle$, $|1011\rangle$, $|1100\rangle$, $|1101\rangle$ and $|1111\rangle$.
Using Eq.(\ref{TH2}) to compute the negativity, we obtain
\begin{eqnarray}\label{TH55}
N^{CL_{4}}_{A(BCD)}=1.
\end{eqnarray}
Exchange symmetry between $A$ and $B$ gives $N^{CL_4}_{B(ACD)}=1$ as well. These two negativities are independent of $T_H$ and saturate the upper bound in Eq.~(\ref{TH1}), establishing freezing of maximal bipartite negativity.

This invariance follows from a factorization of the state and local channel. Define
\[
\alpha=(1+e^{-\omega/T_H})^{-1/2},\qquad
\beta=(1+e^{\omega/T_H})^{-1/2},\qquad
\alpha^2+\beta^2=1.
\]
Tracing over the interior partner in Eq.~(\ref{w20}) yields the
reduced channel in Kraus form~\cite{Kraus1971},
\[
\mathcal E_{T_H}(X)=K_0XK_0^\dagger+K_1XK_1^\dagger,
\]
where
\[
K_0=\alpha|0\rangle\langle0|+|1\rangle\langle1|,
\qquad K_1=\beta|1\rangle\langle0|.
\]
These operators satisfy
\[
K_0^\dagger K_0+K_1^\dagger K_1=I_D.
\]
Writing $|\psi\rangle=|\Psi^{CL_4}_{ABCD}\rangle$, the accessible state is
\[
\rho(T_H)=(\mathcal I_{ABC}\otimes\mathcal E_{T_H}^{(D)})
(|\psi\rangle\langle\psi|),
\]
where $\mathcal I$ denotes the identity map on the indicated subsystems.
The controlled-phase unitary $U=CZ_{BC}$, with $CZ=\operatorname{diag}(1,1,1,-1)$, gives
\[
U|\psi\rangle=|\Phi^+\rangle_{AB}\otimes|\Phi^+\rangle_{CD},
\qquad |\Phi^+\rangle=\frac{|00\rangle+|11\rangle}{\sqrt2}.
\]
As $U$ acts on $BC$, it commutes with the map on $D$, and hence
\[
U\rho(T_H)U^\dagger
=|\Phi^+\rangle\langle\Phi^+|_{AB}\otimes\sigma_{CD}(T_H),
\]
where
\[
\sigma_{CD}(T_H)=(\mathcal I_C\otimes\mathcal E_{T_H}^{(D)})
(|\Phi^+\rangle\langle\Phi^+|_{CD}).
\]
For $A|BCD$, $U$ acts entirely within the complementary subsystem. Unitary invariance and multiplicativity of the trace norm therefore imply
\[
\|\rho(T_H)^{T_A}\|
=\|(|\Phi^+\rangle\langle\Phi^+|_{AB})^{T_A}\|
\|\sigma_{CD}(T_H)\|=2,
\]
where $\|\sigma_{CD}(T_H)\|=1$. This proves $N^{CL_4}_{A(BCD)}=1$, and the preserved $A\leftrightarrow B$ symmetry gives the same result for $B|ACD$. The unaffected Bell factor also makes the proof valid for any completely positive trace-preserving map confined to $D$, within the same mode tensor product.

\begin{figure}
\begin{minipage}[t]{\linewidth}
\centering
\includegraphics[width=3.5in,height=6.5cm]{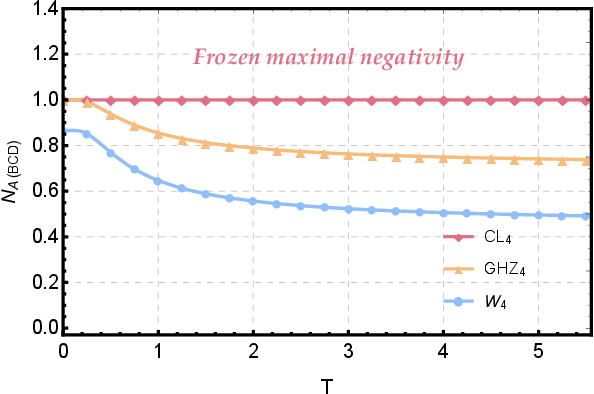}
\label{fig1}
\end{minipage}%
\caption{The negativity $N_{A(BCD)}$ of the $CL_4$, $GHZ_4$, and $W_4$ states versus $T\equiv T_H$, at fixed Schwarzschild Killing frequency $\omega=1$, with the Hawking map acting on $D$.}
\label{Fig1}
\end{figure}

Fig.\ref{Fig1} compares $N_{A(BCD)}$ for the $CL_4$, $GHZ_4$, and $W_4$ states at fixed $\omega=1$. The cluster-state negativity remains maximal throughout the temperature range, whereas the GHZ- and W-state negativities decrease towards finite limits. The comparison-state calculations are given in Appendices~A and~B. For the cluster state, the Bell-factor decomposition explains why the specified bipartite negativity remains invariant even though the reduced density matrix depends on temperature. The freezing thus reflects the structure of the state relative to the chosen bipartition and affected mode.

We now compute the entanglement for the remaining single-qubit-versus-rest bipartitions by taking the partial transpose with respect to modes $C$ and $D$. The resulting partially transposed density matrices can be expressed as
\begin{eqnarray}\label{TH333}
\rho^{CL_{4}({T_{C}})}_{ABCD}=
\frac{1}{4}
 \left(\!\!\begin{array}{cccccccc}
\frac{1}{e^{-\frac{\omega}{T}}+1} & 0 & 0 & {\hskip 3pt}0{\hskip 3pt} & \frac{1}{e^{-\frac{\omega}{T}}+1} & 0 & 0 & {\hskip 3pt}0 \\
0 & \frac{1}{e^{\frac{\omega}{T}}+1} & \frac{1}{\sqrt{e^{-\frac{\omega}{T}}+1}} & {\hskip 3pt}0{\hskip 3pt} & 0 & \frac{1}{e^{\frac{\omega}{T}}+1} & \frac{1}{\sqrt{e^{-\frac{\omega}{T}}+1}} & {\hskip 3pt}0 \\
0 & \frac{1}{\sqrt{e^{-\frac{\omega}{T}}+1}} & 0 & {\hskip 3pt}0{\hskip 3pt} & 0 & -\frac{1}{\sqrt{e^{-\frac{\omega}{T}}+1}} & 0 & {\hskip 3pt}0 \\
0 & 0 & 0 & {\hskip 3pt}1{\hskip 3pt} & 0 & 0 & 0 & {\hskip 3pt}-1 \\
\frac{1}{e^{-\frac{\omega}{T}}+1} & 0 & 0 & {\hskip 3pt}0{\hskip 3pt} & \frac{1}{e^{-\frac{\omega}{T}}+1} & 0 & 0 & {\hskip 3pt}0 \\
0 & \frac{1}{e^{\frac{\omega}{T}}+1} & -\frac{1}{\sqrt{e^{-\frac{\omega}{T}}+1}} & {\hskip 3pt}0{\hskip 3pt} & 0 & \frac{1}{e^{\frac{\omega}{T}}+1}& -\frac{1}{\sqrt{e^{-\frac{\omega}{T}}+1}}& {\hskip 3pt}0 \\
0 & \frac{1}{\sqrt{e^{-\frac{\omega}{T}}+1}} & 0 & 0 & {\hskip 3pt}0{\hskip 3pt} & -\frac{1}{\sqrt{e^{-\frac{\omega}{T}}+1}} & 0 & {\hskip 3pt}0 \\
0 & 0 & 0 & {\hskip 3pt}-1{\hskip 3pt} & 0 & 0 & 0 & {\hskip 3pt}1
\end{array}\!\!\right),
\end{eqnarray}

\begin{eqnarray}\label{TH4}
\rho^{CL_{4}({T_{D}})}_{ABCD}=
\frac{1}{4}
 \left(\!\!\begin{array}{cccccccc}
\frac{1}{e^{-\frac{\omega}{T}}+1} & 0 & 0 & {\hskip 3pt}0{\hskip 3pt} & \frac{1}{e^{-\frac{\omega}{T}}+1} & 0 & 0 & {\hskip 3pt}0 \\
0 & \frac{1}{e^{\frac{\omega}{T}}+1} & \frac{1}{\sqrt{e^{-\frac{\omega}{T}}+1}} & {\hskip 3pt}0{\hskip 3pt} & 0 & \frac{1}{e^{\frac{\omega}{T}}+1} & -\frac{1}{\sqrt{e^{-\frac{\omega}{T}}+1}} & {\hskip 3pt}0 \\
0 & \frac{1}{\sqrt{e^{-\frac{\omega}{T}}+1}} & 0 & {\hskip 3pt}0{\hskip 3pt} & 0 & \frac{1}{\sqrt{e^{-\frac{\omega}{T}}+1}} & 0 & {\hskip 3pt}0 \\
0 & 0 & 0 & {\hskip 3pt}1{\hskip 3pt} & 0 & 0 & 0 & {\hskip 3pt}-1 \\
\frac{1}{e^{-\frac{\omega}{T}}+1} & 0 & 0 & {\hskip 3pt}0{\hskip 3pt} & \frac{1}{e^{-\frac{\omega}{T}}+1} & 0 & 0 & {\hskip 3pt}0 \\
0 & \frac{1}{e^{\frac{\omega}{T}}+1} & \frac{1}{\sqrt{e^{-\frac{\omega}{T}}+1}} & {\hskip 3pt}0{\hskip 3pt} & 0 & \frac{1}{e^{\frac{\omega}{T}}+1} &-\frac{1}{\sqrt{e^{-\frac{\omega}{T}}+1}} & {\hskip 3pt}0 \\
0 & -\frac{1}{\sqrt{e^{-\frac{\omega}{T}}+1}} & 0 & 0 & {\hskip 3pt}0{\hskip 3pt} & -\frac{1}{\sqrt{e^{-\frac{\omega}{T}}+1}} & 0 & {\hskip 3pt}0 \\
0 & 0 & 0 & {\hskip 3pt}-1{\hskip 3pt} & 0 & 0 & 0 & {\hskip 3pt}1
\end{array}\!\!\right),
\end{eqnarray}
where the matrix $\rho^{{CL_{4}}({T_{C}})}_{ABCD}$ is expressed in the ordered basis $|0000\rangle$, $|0001\rangle$, $|0010\rangle$, $|0011\rangle$, $|1100\rangle$, $|1101\rangle$, $|1110\rangle$ and $|1111\rangle$,
and the matrix $\rho^{{CL_{4}}({T_{D}})}_{ABCD}$ is expressed in the ordered basis $|0000\rangle$, $|0001\rangle$, $|0010\rangle$, $|0011\rangle$, $|1100\rangle$, $|1101\rangle$, $|1110\rangle$ and $|1111\rangle$.
Applying Eq.(\ref{TH2}), the corresponding negativities can be expressed as
\begin{eqnarray}\label{TH555}
N^{CL_{4}}_{C(ABD)}=\max\left\{0,\frac{1}{\sqrt{1+e^{-\frac{\omega}{T}}}}\right\},
\end{eqnarray}
\begin{eqnarray}\label{TH5}
N^{CL_{4}}_{D(ABC)}=\max\left\{0,\frac{1}{2}\left[\sqrt{ \frac{4}{1+e^{-\frac{\omega}{T}}}+\frac{1}{(1+e^{\frac{\omega}{T}})^{2}} }
-\frac{1}{1+e^{\frac{\omega}{T}}}\right] \right\} =\max\left\{0,
\frac{1}{1+e^{-\frac{\omega}{T}}}  \right\}.
\end{eqnarray}

To determine the dependence on the affected mode, we apply the same Hawking map to each mode of the fixed input state in Eq.~(\ref{w23}). Eqs~(\ref{TH555}) and~(\ref{TH5}) simplify to $\alpha$ and $\alpha^2$, respectively. The exchange symmetries of the input state then give the results in Table~\ref{tab:channel-location}.
\begin{table}[t]
\centering
\caption{Single-qubit-versus-rest negativities of the cluster state for different locations of the Hawking map, with $\alpha=(1+e^{-\omega/T_H})^{-1/2}$.}
\label{tab:channel-location}
\begin{tabular}{c|cccc}
\hline
Mode acted on & $N_{A(BCD)}$ & $N_{B(ACD)}$ & $N_{C(ABD)}$ & $N_{D(ABC)}$ \\
\hline
$A$ & $\alpha^2$ & $\alpha$ & $1$ & $1$ \\
$B$ & $\alpha$ & $\alpha^2$ & $1$ & $1$ \\
$C$ & $1$ & $1$ & $\alpha^2$ & $\alpha$ \\
$D$ & $1$ & $1$ & $\alpha$ & $\alpha^2$ \\
\hline
\end{tabular}
\end{table}
The maximal negativities are preserved across $A|BCD$ and $B|ACD$ when the map acts on $C$ or $D$. Moving the map to $A$ or $B$ exchanges the invariant cuts with $C|ABD$ and $D|ABC$. Thus, the location of the channel determines which bipartitions retain their initial maximal negativity for this state.

\begin{figure}
\begin{minipage}[t]{0.5\linewidth}
\centering
\includegraphics[width=3.0in,height=6.24cm]{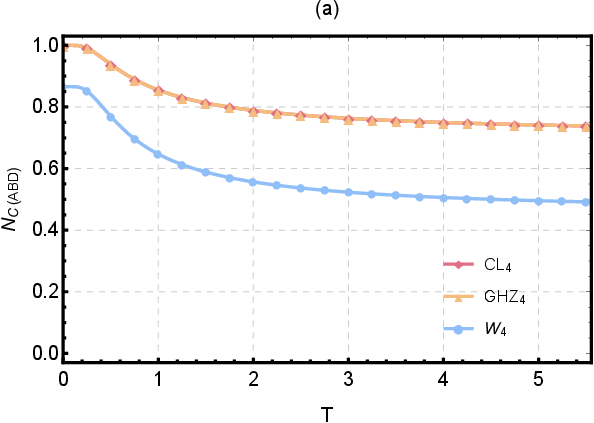}
\label{fig2a}
\end{minipage}%
\begin{minipage}[t]{0.5\linewidth}
\centering
\includegraphics[width=3.0in,height=6.24cm]{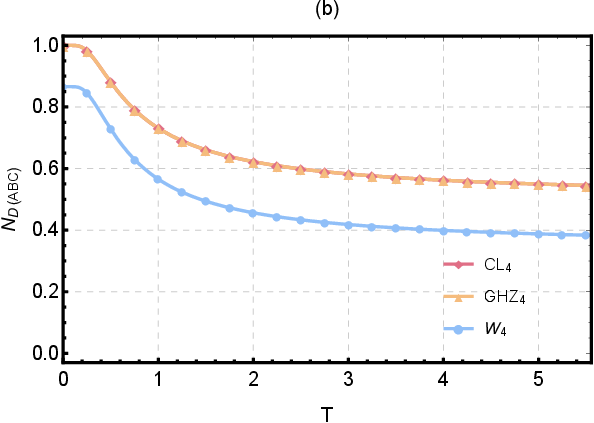}
\label{fig2b}
\end{minipage}%
\caption{The negativities (a) $N_{C(ABD)}$ and (b) $N_{D(ABC)}$ of the $CL_4$, $GHZ_4$, and $W_4$ states versus $T\equiv T_H$, at fixed Schwarzschild Killing frequency $\omega=1$, with the Hawking map acting on $D$.}
\label{Fig2}
\end{figure}

Fig.\ref{Fig2} presents the negativities across $C|ABD$ and $D|ABC$ for the three states, with the Hawking map acting on $D$. The cluster and GHZ states have identical curves, $N_{C(ABD)}=\alpha$ and $N_{D(ABC)}=\alpha^2$, approaching $1/\sqrt2$ and $1/2$, respectively, as $T_H$ increases without bound. For these states, $N_{C(ABD)}>N_{D(ABC)}$ at $T_H>0$, and both tend to one in the zero-temperature limit. The W-state curves lie below them; their initial negativity is $\sqrt3/2$, compared with one for the cluster and GHZ states. The ordering of absolute values therefore reflects both the initial entanglement and its response to the map. Together with Fig.~\ref{Fig1}, these results reveal the coexistence of invariant and temperature-dependent bipartite negativities in the cluster state.

The analysis uses occupation-number qubits associated with selected field modes. Relating this description to localized wave packets or detector qubits requires specifying the preparation, detector trajectory, field coupling, and accessible modes~\cite{SDF7,Bruschi2010}. The Bell-factor proof continues to hold whenever the resulting description preserves the same subsystem tensor product and confines the effective channel to $D$. These conditions provide a starting point for investigating operational realizations of the invariant bipartite entanglement. The high-temperature limit is taken at the level of the reduced map within this fixed-background treatment.

\section{Conclusions}
We have demonstrated exact freezing of maximal bipartite negativity for a four-qubit cluster state in Schwarzschild spacetime within a fermionic single-mode description. When the Hawking transformation acts on $D$, the negativities across $A|BCD$ and $B|ACD$ remain equal to one at all Hawking temperatures, while those across $C|ABD$ and $D|ABC$ decrease towards $1/\sqrt{2}$ and $1/2$, respectively, in the formal high-temperature limit. An analytical factorization exposes an unaffected Bell-state factor, explaining why the specified negativities remain maximal even though the accessible state varies with temperature. The same argument establishes their invariance under any completely positive trace-preserving map confined to $D$ in the chosen mode tensor product.

Applying the Hawking map to different modes determines the corresponding invariant bipartitions. In particular, a map on $A$ or $B$ preserves the negativities across $C|ABD$ and $D|ABC$, demonstrating the dependence of freezing on channel location. Comparisons with GHZ and W states further highlight the role of the input entanglement structure under the same map. Together, these results connect the persistence of maximal bipartite mode entanglement to the joint choice of state, bipartition, and channel location. This analytical mechanism provides a basis for exploring entanglement-preserving mode encodings in relativistic quantum information.

\begin{acknowledgments}
 This work is supported by the National Natural Science Foundation of China (12575056), the Natural
Science Foundation of the Liaoning Scientific Committee (2026-MS-281), and the LiaoNing Revitalization Talents Program (XLYC2503099).
\end{acknowledgments}

%\newpage

\appendix
%\begin{widetext}
\onecolumngrid
\section{Quantum entanglement of the tetrapartite $GHZ$ state in Schwarzschild spacetime }
In this appendix, we analyze the properties of  the tetrapartite $GHZ$ state within the same physical setup as for the $CL_4$ state in Sec.~III (see Fig.\ref{Fig0}).
The initial GHZ state of the four mode subsystems is
\begin{eqnarray}\label{gw23}
|\Psi^{GHZ_4}_{ABCD}\rangle=\frac{1}{\sqrt{2}}(|0_A0_B0_C0_D\rangle+|1_A1_B1_C1_D\rangle).
\end{eqnarray}
To incorporate the Hawking effect on mode $D$, we employ Eq.~(\ref{w20}), which introduces its inaccessible interior partner $\bar D$. The expanded state is
\begin{eqnarray}\label{gw25}
|\Psi^{GHZ_4}_{ABCD{\bar D}}\rangle&=&
\frac{1}{\sqrt{2}}\big( \frac{1}{\sqrt{e^{-\frac{\omega}{T}}+1}} |0\rangle_{A}|0\rangle_{B}|0\rangle_{C}|0\rangle_{D}|0\rangle_{\bar D}+\frac{1}{\sqrt{e^{\frac{\omega}{T}}+1}} |0\rangle_{A}|0\rangle_{B}|0\rangle_{C}|1\rangle_{D}|1\rangle_{\bar D}\nonumber\\&&
+|1\rangle_{A}|1\rangle_{B}|1\rangle_{C}|1\rangle_{D}|0\rangle_{\bar D} \big).
\end{eqnarray}
Tracing out the inaccessible mode $\bar{D}$ yields the accessible reduced density matrix of the four mode subsystems, which in the subspace spanned by $|0000\rangle, |0001\rangle, |1111\rangle$ reads
\begin{eqnarray}\label{gw27}
 \rho^{GHZ_4}_{ABCD}=\frac{1}{2}
 \left(\!\!\begin{array}{ccc}
\frac{1}{e^{-\frac{\omega}{T}}+1} & 0 & \frac{1}{\sqrt{e^{-\frac{\omega}{T}}+1}} \\
0 & \frac{1}{e^{\frac{\omega}{T}}+1} & 0 \\
\frac{1}{\sqrt{e^{-\frac{\omega}{T}}+1}} & 0 & 1
\end{array}\!\!\right).
\end{eqnarray}

We then evaluate the single-qubit-versus-rest negativities of $\rho^{GHZ_4}_{ABCD}$. Taking the partial transpose with respect to modes $A$, $C$ and $D$, we obtain the following matrices
\begin{eqnarray}\label{gTH3}
\rho^{{GHZ_4}({T_{A}})}_{ABCD}=
\frac{1}{2}
 \left(\!\!\begin{array}{ccccc}
\frac{1}{e^{-\frac{\omega}{T}}+1} & 0 & 0 & 0 & {\hskip 6pt}0{\hskip 6pt} \\
0 & \frac{1}{e^{\frac{\omega}{T}}+1} & 0 & 0 & 0 \\
0 & 0 & 0 & \frac{1}{\sqrt{e^{-\frac{\omega}{T}}+1}} & 0 \\
0 & 0 & \frac{1}{\sqrt{e^{-\frac{\omega}{T}}+1}} & 0 & 0 \\
0 & 0 & 0 & 0 & 1
\end{array}\!\!\right),
\end{eqnarray}
\begin{eqnarray}\label{gTH333}
\rho^{{GHZ_4}({T_{C}})}_{ABCD}=
\frac{1}{2}
 \left(\!\!\begin{array}{ccccc}
\frac{1}{e^{-\frac{\omega}{T}}+1} & 0 & 0 & 0 & {\hskip 6pt}0{\hskip 6pt} \\
0 & \frac{1}{e^{\frac{\omega}{T}}+1} & 0 & 0 & 0 \\
0 & 0 & 0 & \frac{1}{\sqrt{e^{-\frac{\omega}{T}}+1}} & 0 \\
0 & 0 & \frac{1}{\sqrt{e^{-\frac{\omega}{T}}+1}} & 0 & 0 \\
0 & 0 & 0 & 0 & 1
\end{array}\!\!\right),
\end{eqnarray}
\begin{eqnarray}\label{gTH4}
\rho^{{GHZ_4}({T_{D}})}_{ABCD}=
\frac{1}{2}
 \left(\!\!\begin{array}{cccc}
\frac{1}{e^{-\frac{\omega}{T}}+1} & 0 & 0 & {\hskip 6pt}0{\hskip 6pt} \\
0 & \frac{1}{e^{\frac{\omega}{T}}+1} & \frac{1}{\sqrt{e^{-\frac{\omega}{T}}+1}} & 0 \\
0 & \frac{1}{\sqrt{e^{-\frac{\omega}{T}}+1}} & 0 & 0 \\
0 & 0 & 0 & 1
\end{array}\!\!\right).
\end{eqnarray}
For the partial transpose with respect to $A$, we use the ordered basis $|0000\rangle$, $|0001\rangle$, $|0111\rangle$, $|1000\rangle$, $|1111\rangle$.
For $C$, the basis is $|0000\rangle$, $|0001\rangle$, $|0010\rangle$, $|1101\rangle$, $|1111\rangle$; for $D$, it is $|0000\rangle$, $|0001\rangle$, $|1110\rangle$, $|1111\rangle$.
Applying Eq.~(\ref{TH2}) gives
\begin{eqnarray}\label{gTH555}
N^{GHZ_4}_{A(BCD)}=N^{GHZ_4}_{C(ABD)}=\max\left\{0,\frac{1}{\sqrt{1+e^{-\frac{\omega}{T}}}}\right\},
\end{eqnarray}

\begin{eqnarray}\label{gTH5}
N^{GHZ_4}_{D(ABC)}=\max\left\{0, \frac{1}{2}
\bigg( \sqrt{ \frac{4}{1+e^{-\frac{\omega}{T}}}+\frac{1}{(1+e^{\frac{\omega}{T}})^{2}} }
-\frac{1}{1+e^{\frac{\omega}{T}}} \bigg) \right\} .
\end{eqnarray}
A key observation is that $N^{GHZ_4}_{C(ABD)}$ and $N^{GHZ_4}_{D(ABC)}$ are mathematically identical to the corresponding results for the $CL_4$ state [Eqs.(\ref{TH555}) and (\ref{TH5})]. This equivalence explains the coincident decay curves of the $CL_4$ and $GHZ_4$ states in Fig.\ref{Fig2}.

\section{Quantum entanglement of the tetrapartite $W$ state in Schwarzschild spacetime }
The initial W state in the same mode encoding is
\begin{eqnarray}\label{ww23}
|\Psi^{W_4}_{ABCD}\rangle=\frac{1}{2}(|0_A0_B0_C1_D\rangle+|0_A0_B1_C0_D\rangle+|0_A1_B0_C0_D\rangle+|1_A0_B0_C0_D\rangle).
\end{eqnarray}
Following the same procedure as for the $CL_4$ and $GHZ_4$ states, we apply the Bogoliubov transformation in
Eq.(\ref{w20}) to mode $D$, which gives the expanded state
\begin{eqnarray}\label{ww25}
|\Psi^{W_4}_{ABCD{\bar D}}\rangle&=&
\frac{1}{2\sqrt{e^{-\frac{\omega}{T}}+1}} \big(|0\rangle_{A}|0\rangle_{B}|1\rangle_{C}|0\rangle_{D}|0\rangle_{\bar D}+|0\rangle_{A}|1\rangle_{B}|0\rangle_{C}|0\rangle_{D}|0\rangle_{\bar D} \nonumber\\&&+|1\rangle_{A}|0\rangle_{B}|0\rangle_{C}|0\rangle_{D}|0\rangle_{\bar D} \big)
+\frac{1}{2\sqrt{e^{\frac{\omega}{T}}+1}} \big(|0\rangle_{A}|0\rangle_{B}|1\rangle_{C}|1\rangle_{D}|1\rangle_{\bar D}\nonumber\\&&
+|0\rangle_{A}|1\rangle_{B}|0\rangle_{C}|1\rangle_{D}|1\rangle_{\bar D} +|1\rangle_{A}|0\rangle_{B}|0\rangle_{C}|1\rangle_{D}|1\rangle_{\bar D} \big)\nonumber\\&&
+\frac{1}{2} |0\rangle_{A}|0\rangle_{B}|0\rangle_{C}|1\rangle_{D}|0\rangle_{\bar D}.
\end{eqnarray}
Tracing out $\bar{D}$ yields the reduced density matrix for the four accessible modes. In the ordered basis $|0001\rangle$, $|0010\rangle$, $|0011\rangle$, $|0100\rangle$, $|0101\rangle$, $|1000\rangle$, $|1001\rangle$, the matrix reads
\begin{eqnarray}\label{ww27}
 \rho^{W_4}_{ABCD}=\frac{1}{4}
 \left(\!\!\begin{array}{ccccccc}
1 & \frac{1}{\sqrt{e^{-\frac{\omega}{T}}+1}} & 0 & \frac{1}{\sqrt{e^{-\frac{\omega}{T}}+1}} & 0 & \frac{1}{\sqrt{e^{-\frac{\omega}{T}}+1}} & 0 \\
\frac{1}{\sqrt{e^{-\frac{\omega}{T}}+1}} & \frac{1}{e^{-\frac{\omega}{T}}+1} & 0 & \frac{1}{e^{-\frac{\omega}{T}}+1} & 0 & \frac{1}{e^{-\frac{\omega}{T}}+1} & 0 \\
0 & 0 & \frac{1}{e^{\frac{\omega}{T}}+1} & 0 & \frac{1}{e^{\frac{\omega}{T}}+1} & 0 &\frac{1}{e^{\frac{\omega}{T}}+1} \\
\frac{1}{\sqrt{e^{-\frac{\omega}{T}}+1}} & \frac{1}{e^{-\frac{\omega}{T}}+1} & 0 & \frac{1}{e^{-\frac{\omega}{T}}+1} & 0 & \frac{1}{e^{-\frac{\omega}{T}}+1} & 0 \\
0 & 0 & \frac{1}{e^{\frac{\omega}{T}}+1} & 0 & \frac{1}{e^{\frac{\omega}{T}}+1} & 0 & \frac{1}{e^{\frac{\omega}{T}}+1} \\
\frac{1}{\sqrt{e^{-\frac{\omega}{T}}+1}} & \frac{1}{e^{-\frac{\omega}{T}}+1} & 0 & \frac{1}{e^{-\frac{\omega}{T}}+1} & 0 & \frac{1}{e^{-\frac{\omega}{T}}+1} & 0 \\
0 & 0 & \frac{1}{e^{\frac{\omega}{T}}+1} & 0 & \frac{1}{e^{\frac{\omega}{T}}+1} & 0 & \frac{1}{e^{\frac{\omega}{T}}+1}
\end{array}\!\!\right).
\end{eqnarray}

Next we compute the single-qubit-versus-rest negativities of the W state by performing the partial transposition with respect to modes $A$, $C$ and $D$. The resulting matrices can be expressed as
\begin{eqnarray}\label{wTH3}
\scalebox{0.8}{$
\rho^{{W_4}({T_{A}})}_{ABCD}=
\frac{1}{4}
 \left(\!\!\begin{array}{cccccccccccc}
0 & 0 & 0 & 0 & 0 & 0 & 0 & \frac{1}{\sqrt{e^{-\frac{\omega}{T}}+1}} & \frac{1}{{e^{-\frac{\omega}{T}}+1}} & 0 & \frac{1}{{e^{-\frac{\omega}{T}}+1}} & 0 \\
0 & 1 & \frac{1}{\sqrt{e^{-\frac{\omega}{T}}+1}} & 0 & \frac{1}{\sqrt{e^{-\frac{\omega}{T}}+1}} & 0 & 0 & 0 & 0 & \frac{1}{{e^{\frac{\omega}{T}}+1}} & 0 & \frac{1}{{e^{\frac{\omega}{T}}+1}} \\
0 & \frac{1}{\sqrt{e^{-\frac{\omega}{T}}+1}} & \frac{1}{e^{-\frac{\omega}{T}}+1} & 0 & \frac{1}{e^{-\frac{\omega}{T}}+1} & 0 & 0 & 0 & 0 & 0 & 0 & 0 \\
0 & 0 & 0 &\frac{1}{e^{\frac{\omega}{T}}+1} & 0 & \frac{1}{e^{\frac{\omega}{T}}+1} & 0 & 0 & 0 & 0 & 0 & 0 \\
0 & \frac{1}{\sqrt{e^{-\frac{\omega}{T}}+1}} & \frac{1}{e^{-\frac{\omega}{T}}+1} & 0 & \frac{1}{e^{-\frac{\omega}{T}}+1} & 0 & 0 & 0 & 0 & 0 & 0 & 0 \\
0 & 0 & 0 & \frac{1}{e^{\frac{\omega}{T}}+1} & 0 & \frac{1}{e^{\frac{\omega}{T}}+1} & 0 & 0 & 0 & 0 & 0 & 0 \\
0 & 0 & 0 & 0 & 0 & 0 & \frac{1}{e^{-\frac{\omega}{T}}+1} & 0 & 0 & 0 & 0 & 0 \\
\frac{1}{\sqrt{e^{-\frac{\omega}{T}}+1}} & 0 & 0 & 0 & 0 & 0 & 0 & \frac{1}{e^{\frac{\omega}{T}}+1} & 0 & 0 & 0 & 0 \\
\frac{1}{e^{-\frac{\omega}{T}}+1} & 0 & 0 & 0 & 0 & 0 & 0 & 0 & 0 & 0 & 0 & 0 \\
0 & \frac{1}{e^{\frac{\omega}{T}}+1} & 0 & 0 & 0 & 0 & 0 & 0 & 0 & 0 & 0 & 0 \\
\frac{1}{e^{-\frac{\omega}{T}}+1} & 0 & 0 & 0 & 0 & 0 & 0 & 0 & 0 & 0 & 0 & 0 \\
0 & \frac{1}{e^{\frac{\omega}{T}}+1} & 0 & 0 & 0 & 0 & 0 & 0 & 0 & 0 & 0 & 0 \\
\end{array}\!\!\right)$},
\end{eqnarray}
\begin{eqnarray}\label{wTH333}
\scalebox{0.8}{$
\rho^{{W_4}({T_{C}})}_{ABCD}=
\frac{1}{4}
 \left(\!\!\begin{array}{cccccccccccc}
0 & 0 & 0 & \frac{1}{\sqrt{e^{-\frac{\omega}{T}}+1}} & 0 & 0 & \frac{1}{e^{-\frac{\omega}{T}}+1} & 0 & 0 & 0 & \frac{1}{{e^{-\frac{\omega}{T}}+1}} & 0 \\
0 & 1 & 0 & 0 & \frac{1}{\sqrt{e^{-\frac{\omega}{T}}+1}} & 0 & 0 & \frac{1}{e^{\frac{\omega}{T}}+1} & \frac{1}{\sqrt{e^{-\frac{\omega}{T}}+1}}& 0 & 0 & \frac{1}{{e^{\frac{\omega}{T}}+1}} \\
0 & 0 & \frac{1}{e^{-\frac{\omega}{T}}+1} & 0 & 0 & 0 & 0 & 0 & 0 & 0 & 0 & 0 \\
\frac{1}{\sqrt{e^{-\frac{\omega}{T}}+1}} & 0 & 0 & \frac{1}{e^{\frac{\omega}{T}}+1} & 0 & 0 & 0 & 0 & 0 & 0 & 0 & 0 \\
0 & \frac{1}{\sqrt{e^{-\frac{\omega}{T}}+1}} & 0 & 0 & \frac{1}{e^{-\frac{\omega}{T}}+1} & 0 & 0 & 0 & \frac{1}{e^{-\frac{\omega}{T}}+1} & 0 & 0 & 0 \\
0 & 0& 0 & 0 & 0 & \frac{1}{e^{\frac{\omega}{T}}+1} & 0 & 0 & 0 & \frac{1}{e^{\frac{\omega}{T}}+1} & 0 & 0 \\
\frac{1}{e^{-\frac{\omega}{T}}+1} & 0 & 0 & 0 & 0 & 0 & 0 & 0 & 0 & 0 & 0 & 0 \\
0 & \frac{1}{e^{\frac{\omega}{T}}+1} & 0 & 0 & 0 & 0 & 0 & 0 & 0 & 0 & 0 & 0 \\
0 & \frac{1}{\sqrt{e^{-\frac{\omega}{T}}+1}} & 0 & 0 & \frac{1}{e^{-\frac{\omega}{T}}+1} & 0 & 0 & 0 & \frac{1}{e^{-\frac{\omega}{T}}+1} & 0 & 0 & 0 \\
0 & 0 & 0 & 0 & 0 & \frac{1}{e^{\frac{\omega}{T}}+1} & 0 & 0 & 0 & \frac{1}{e^{\frac{\omega}{T}}+1} & 0 & 0 \\
\frac{1}{e^{-\frac{\omega}{T}}+1} & 0 & 0 & 0 & 0 & 0 & 0 & 0 & 0 & 0 & 0 & 0 \\
0 & \frac{1}{e^{\frac{\omega}{T}}+1} & 0 & 0 & 0 & 0 & 0 & 0 & 0 & 0 & 0 & 0
\end{array}\!\!\right),
$}
\end{eqnarray}
\begin{eqnarray}\label{wTH4}
\rho^{{W_4}({T_{D}})}_{ABCD}=
 \frac{1}{4}
 \left(\!\!\begin{array}{cccccccc}
0 & {\hskip 6pt}0{\hskip 6pt} & 0 & \frac{1}{\sqrt{e^{-\frac{\omega}{T}}+1}} & 0 & \frac{1}{\sqrt{e^{-\frac{\omega}{T}}+1}} & 0 & \frac{1}{\sqrt{e^{-\frac{\omega}{T}}+1}} \\
0 & 1 & 0 & 0 & 0 & 0 & 0 & 0 \\
0 & 0 & \frac{1}{e^{-\frac{\omega}{T}}+1} & 0 & \frac{1}{e^{-\frac{\omega}{T}}+1} & 0 & \frac{1}{e^{-\frac{\omega}{T}}+1} & 0 \\
\frac{1}{\sqrt{e^{-\frac{\omega}{T}}+1}} & 0 & 0 & \frac{1}{e^{\frac{\omega}{T}}+1} & 0 & \frac{1}{e^{\frac{\omega}{T}}+1} & 0 & \frac{1}{e^{\frac{\omega}{T}}+1} \\
0 & 0 & \frac{1}{e^{-\frac{\omega}{T}}+1} & 0 & \frac{1}{e^{-\frac{\omega}{T}}+1} & 0 & \frac{1}{e^{-\frac{\omega}{T}}+1} & 0 \\
\frac{1}{\sqrt{e^{-\frac{\omega}{T}}+1}} & 0 & 0 & \frac{1}{e^{\frac{\omega}{T}}+1} & 0 & \frac{1}{e^{\frac{\omega}{T}}+1} & 0 & \frac{1}{e^{\frac{\omega}{T}}+1} \\
0 & 0 & \frac{1}{e^{-\frac{\omega}{T}}+1} & 0 & \frac{1}{e^{-\frac{\omega}{T}}+1} & 0 & \frac{1}{e^{-\frac{\omega}{T}}+1} & 0 \\
\frac{1}{\sqrt{e^{-\frac{\omega}{T}}+1}} & 0 & 0 & \frac{1}{e^{\frac{\omega}{T}}+1} & 0 & \frac{1}{e^{\frac{\omega}{T}}+1} & 0 & \frac{1}{e^{\frac{\omega}{T}}+1}
\end{array}\!\!\right),
\end{eqnarray}
where $\rho^{{W_4}({T_{A}})}_{ABCD}$ is expressed in the ordered basis $|0000\rangle$, $|0001\rangle$, $|0010\rangle$, $|0011\rangle$, $|0100\rangle$, $|0101\rangle$, $|1000\rangle$, $|1001\rangle$, $|1010\rangle$, $|1011\rangle$, $|1100\rangle$, $|1101\rangle$;
$\rho^{{W_4}({T_{C}})}_{ABCD}$ in $|0000\rangle$, $|0001\rangle$, $|0010\rangle$, $|0011\rangle$, $|0100\rangle$, $|0101\rangle$, $|0110\rangle$, $|0111\rangle$, $|1000\rangle$, $|1001\rangle$, $|1010\rangle$, $|1011\rangle$;
and $\rho^{{W_4}({T_{D}})}_{ABCD}$ in $|0000\rangle$, $|0001\rangle$, $|0010\rangle$, $|0011\rangle$, $|0100\rangle$, $|0101\rangle$, $|1000\rangle$, $|1001\rangle$.
Applying Eq.(\ref{TH2}) to these matrices yields the corresponding negativities. Due to the complexity of the expressions for $N^{W_4}_{A(BCD)}$ and $N^{W_4}_{C(ABD)}$, we refrain from presenting them explicitly here. For the negativity $N^{W_4}_{D(ABC)}$, we obtain the following compact expression
\begin{eqnarray}\label{wTH5}
N^{W_4}_{D(ABC)}=\max\left\{0, \frac{1}{4}
\bigg( \sqrt{ \frac{12}{1+e^{-\frac{\omega}{T}}}+\frac{9}{(1+e^{\frac{\omega}{T}})^{2}} }
-\frac{3}{1+e^{\frac{\omega}{T}}} \bigg) \right\} .
\end{eqnarray}

\end{document}